\documentclass[aps, prl, twocolumn,nofootinbib,superscriptaddress]{revtex4-2}

\usepackage[mathcal]{euscript}
\usepackage{amsmath}
\usepackage{textcomp, gensymb}
\usepackage{chemformula}
\usepackage[version=4]{mhchem} 
\usepackage{amssymb}
\usepackage[dvipsnames]{xcolor}

\usepackage[normalem]{ulem} 
\usepackage{xr}
\usepackage[left]{lineno} 
\usepackage{hyperref}
\usepackage{dsfont}
\usepackage{textcomp}
\usepackage{gensymb}

\usepackage{physics} 

\def \MBT {MnBi\textsubscript{2}Te\textsubscript{4}~}
\def \MBTT {MnBi\textsubscript{2}Te\textsubscript{4}}
\def \LK {Lifshitz-Kosevich~}
\def \SdH {Shubnikov-de-Haas~}

\makeatletter
\DeclareUnicodeCharacter{2212}{\textendash}
\newcommand*{\addFileDependency}[1]{

\typeout{(#1)}
\@addtofilelist{#1}
\newcommand{\beq}{\begin{equation}}
	\newcommand{\eeq}{\end{equation}}

\IfFileExists{#1}{}{\typeout{No file #1.}}
}\makeatother

\begin{document}
\title{Transport evidence of surface states in magnetic topological insulator MnBi\textsubscript{2}Te\textsubscript{4}}

\author{Michael Wissmann}
\affiliation{Leibniz Institute for Solid State and Materials Research,
IFW Dresden, Helmholtzstrasse 20, 01069 Dresden, Germany}
\affiliation{Université Grenoble Alpes, CNRS, CEA, Grenoble-INP, Spintec, 38000 Grenoble, France}

\author{Romain Giraud}
\affiliation{Leibniz Institute for Solid State and Materials Research,
IFW Dresden, Helmholtzstrasse 20, 01069 Dresden, Germany}
\affiliation{Université Grenoble Alpes, CNRS, CEA, Grenoble-INP, Spintec, 38000 Grenoble, France}

\author{B\"orge Mehlhorn}
\affiliation{Leibniz Institute for Solid State and Materials Research,
IFW Dresden, Helmholtzstrasse 20, 01069 Dresden, Germany}
\affiliation{W\"{u}rzburg-Dresden Cluster of Excellence ct.qmat, 01062 Dresden, Germany}

\author{Maxime Leroux}
\affiliation{CNRS, Laboratoire National des Champs Magnétiques Intenses, Université Grenoble-Alpes, Université Toulouse 3, INSA-Toulouse, EMFL, 31400 Toulouse, France}

\author{Mathieu Pierre}
\affiliation{CNRS, Laboratoire National des Champs Magnétiques Intenses, Université Grenoble-Alpes, Université Toulouse 3, INSA-Toulouse, EMFL, 31400 Toulouse, France}

\author{Michel Goiran}
\affiliation{CNRS, Laboratoire National des Champs Magnétiques Intenses, Université Grenoble-Alpes, Université Toulouse 3, INSA-Toulouse, EMFL, 31400 Toulouse, France}

\author{Walter Escoffier}
\affiliation{CNRS, Laboratoire National des Champs Magnétiques Intenses, Université Grenoble-Alpes, Université Toulouse 3, INSA-Toulouse, EMFL, 31400 Toulouse, France}

\author{Bernd B\"{u}chner}
\affiliation{Leibniz Institute for Solid State and Materials Research,
IFW Dresden, Helmholtzstrasse 20, 01069 Dresden, Germany}
\affiliation{W\"{u}rzburg-Dresden Cluster of Excellence ct.qmat, 01062 Dresden, Germany}
\affiliation{Department of Physics, TU Dresden, 01062 Dresden, Germany}

\author{Anna Isaeva}
\affiliation{Institute of Physics, University of Amsterdam, 1098 XH Amsterdam, The Netherlands}
\affiliation{Faculty of Physics, Technical University of Dortmund, 44227 Dortmund, Germany and\\ Research Center "Future Energy Materials and Systems" (RC FEMS), 44227, Dortmund, Germany}

\author{Joseph Dufouleur}
\affiliation{Leibniz Institute for Solid State and Materials Research,
IFW Dresden, Helmholtzstrasse 20, 01069 Dresden, Germany}
\affiliation{W\"{u}rzburg-Dresden Cluster of Excellence ct.qmat, 01062 Dresden, Germany}

\author{Louis Veyrat}
\affiliation{CNRS, Laboratoire National des Champs Magnétiques Intenses, Université Grenoble-Alpes, Université Toulouse 3, INSA-Toulouse, EMFL, 31400 Toulouse, France}
	
\date{\today}

\begin{abstract}
Magnetic topological insulators can host chiral 1D edge channels at zero magnetic field, when a magnetic gap opens at the Dirac point in the band structure of 2D topological surface states, leading to the quantum anomalous Hall effect in ultra-thin nanostructures. 
For thicker nanostructures, quantization is severely reduced by the co-existence of edge states with other quasi-particles, usually considered as bulk states. Yet, surface states also exist above the magnetic gap, but it remains difficult to identify electronic subbands by electrical measurements due to strong disorder. Here we unveil surface states in \MBT nanostructures, using magneto-transport in very-high magnetic fields up to 55 T, giving evidence of Shubnikov-de-Haas oscillations above 40 T. A detailed analysis confirms the 2D nature of these quantum oscillations, thus establishing an alternative method to photoemission spectroscopy for the study of topological surface states in magnetic topological insulators, using Landau level spectroscopy.
\end{abstract}
\maketitle

Topological insulators have metallic interface states at their boundaries, which can be either surface states for 3D systems, or edge states for 2D systems \cite{Thouless1982,Hasan2010,Tokura2019}. These states are of great interest for applications in spintronics due to their helical spin textures with spin-momentum locking \cite{Mellnik2014,Rojas2016}, and they are weakly scattered by disorder. In non-magnetic 3D topological insulators, the anisotropic scattering of topological surface states (TSS) results in an increase of their backscattering length, as compared to other quasi-particles, which is associated to the enhanced transport mobility of TSS with respect to the mobility of bulk carriers \cite{Dufouleur2016} and which is at the origin of quasi-ballistic transport in narrow quantum wires \cite{Dufouleur2013,Dufouleur2017,Dufouleur2018}. 
In magnetic topological insulators, a small non-trivial gap opens in the 2D band structure of surface states due to the exchange field, and 1D ballistic edge states contribute to the quantum anomalous Hall effect \cite{Chang2023}. However, the quantization of the transverse resistance in ultra-thin structures, important for quantum metrology \cite{Patel2024,Poirier2025}, can be severely reduced by bulk carriers \cite{Rosen2022}. 
Actually, in thicker nanostructures, a major limitation to have quantized transport properties could rather come from surface electronic states, which exist above the magnetic gap but within the bulk-band gap. However, due to strong disorder in magnetic TIs, it remains difficult to evidence the nature of electronic subbands by electrical measurements. 

One convenient way to individually access the doping and energy levels of the surface and bulk electronic populations is to perform Landau level spectroscopy. The measurement of Shubnikov-de-Haas oscillations (SdHO) of the resistance under external magnetic field can give access to the carrier density, the transport and the quantum mobilities, the effective mass and the chemical potential of different charge carrier populations, whose contributions to the total conductance can be disentangled so. However, this technique requires magnetic fields such that $\mu B >1$, $\mu$ being the charge carrier mobility, different for each band. While such quantum oscillations have been studied in Sb-doped \MBTT,\cite{Jiang2021,Lee2021}, they were not observed in pure \MBT in the many magneto-transport studies up to 15~T\cite{Cui2019, Lee2019, Chen2019a, Bac2022, Martini2023, Cao2024}, except for the report of small resistance oscillations \cite{Lei2022}, which are not attributed to surface states, but to Weyl physics. This absence of SdHO in pure \MBT is most probably due to a low mobility caused by the strong disorder in the bulk, attributed mainly to Mn/Bi antisites \cite{Otrokov2019,Yan2019,Liu2020,Li2024b,Souchay2019,Zeugner2019}, thus requiring very high magnetic fields to observe SdHO.

\begin{figure}[h]
\centering
\includegraphics[width=0.85\columnwidth]{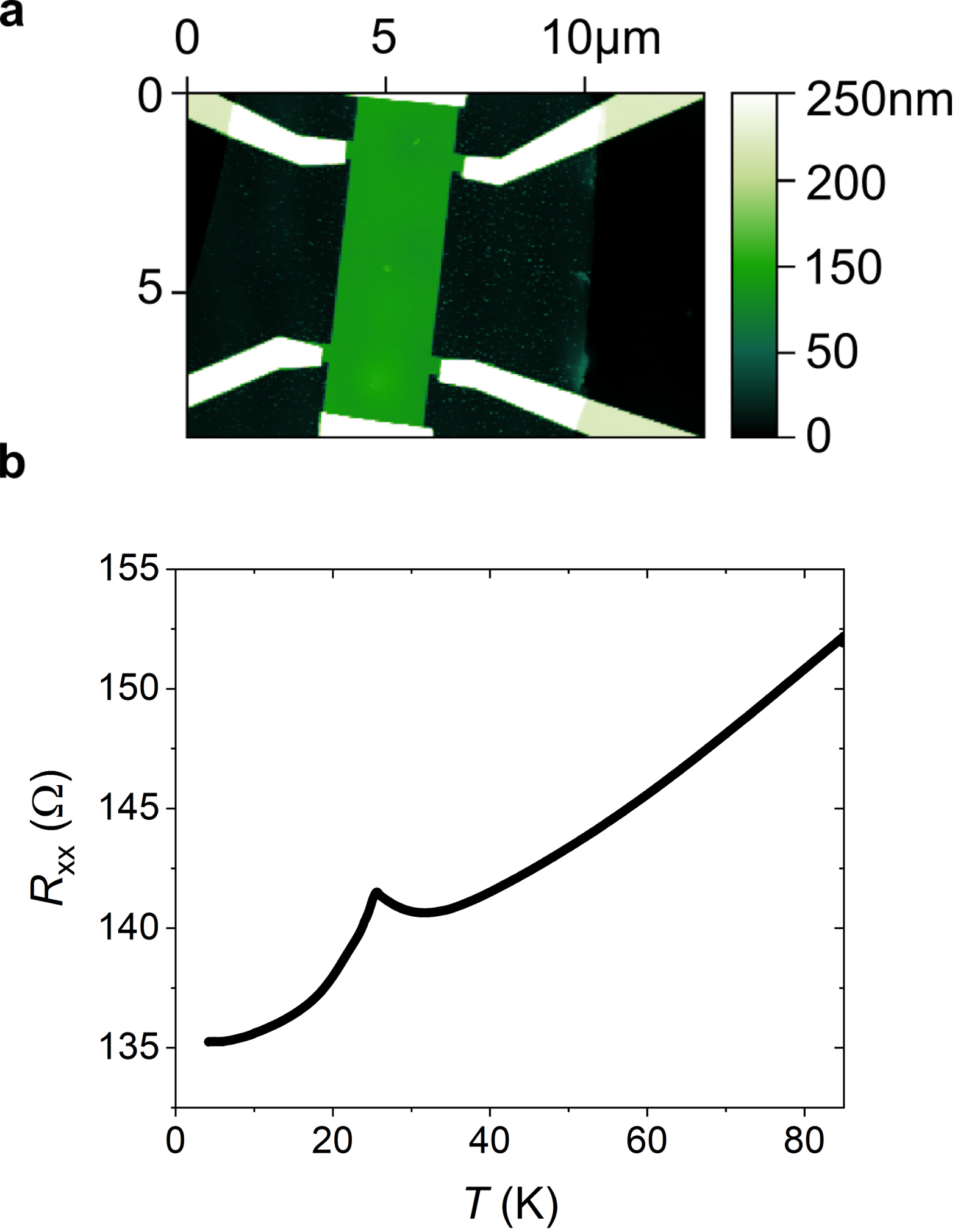}
\caption{
\textbf{Device MBT2: Hall-bar geometry and metallic behavior. a}: 
Atomic force microscopy picture of an etched \MBT nanostructure, with thickness 91~nm and roughness of about 1~nm. The original shape of the \MBT exfoliated flake can still be faintly seen below the contacts.
\textbf{b}: temperature dependence of the longitudinal resistance in zero magnetic field. The antiferromagnetic transition induces a resistance peak at 25.5~K.
}
\label{Fig:charac}
\end{figure}

Here, we report magnetotransport measurements in Hallbar nanostructures of exfoliated \MBT up to very high magnetic fields (55~T) for the first time, revealing the clear signature of SdHO above 40~T. From the detailed study of their temperature and angular dependence, we identify the 2D nature of the electronic band responsible for these SdHO. By comparing the band-specific SdHO carrier density to the average carrier density extracted from the Hall effect, we determine the band bending of the bulk band at the top surface of the nanostructure.
This study provides the first transport evidence of surface states in magnetic topological insulators.
\\ \\
We prepare our samples by exfoliating flakes from high-quality MnBi$_2$Te$_4$ single crystals grown by a controlled protocol developed by us in \cite{Otrokov2019,Zeugner2019}. The products of all growth experiments are routinely checked by powder X-ray diffraction (X'Pert Pro diffractometer (PANalytical), Bragg-Brentano geometry with a variable divergence slit, a curved Ge(111) monochromator, Cu-K$\alpha_1$ radiation ($\lambda = 154.056\,\text{pm}$). Phase analysis of the growth batches is performed by Le Bail profile-fitting method in JANA2006, and shows that the samples are multi-phase as expected for the incongruently melting MnBi$_2$Te$_4$, as explained in full detail in Ref. \cite{Zeugner2019}. The growth batches contain well-shaped platelet-like single crystals that are mechanically extracted and confirmed as MnBi$_2$Te$_4$ by energy-dispersive X-ray spectroscopy (see Fig. S8 in the SI). Their average composition is Te 59.4(7) at. \%, Bi 28.1(8) at. \%, Mn 12.5(4) at. \% in accordance with \cite{Zeugner2019}. These exact verified crystals are then used for further exfoliation.\\
 The exfoliated flakes were then patterned using e-beam lithography to make ohmic Ti/Au contacts and a \ce{TiO2} hard mask to etch a Hall bar. We investigated two devices that yielded very similar results, as shown in Supplementary Information. An atomic force microscopy picture of one such device is presented in Fig.\ref{Fig:charac}a. The flake’s thickness is 91~nm, with a low surface roughness of about 1~nm. The temperature dependence of the longitudinal resistance shows a dirty-metal behavior (low residual resistance ratio), with a pronounced resistance peak at 25.5~K at the antiferromagnetic N\'eel transition temperature $T_\mathrm{N}$.

\begin{figure}[t]
\centering
\includegraphics[width=0.85\columnwidth]{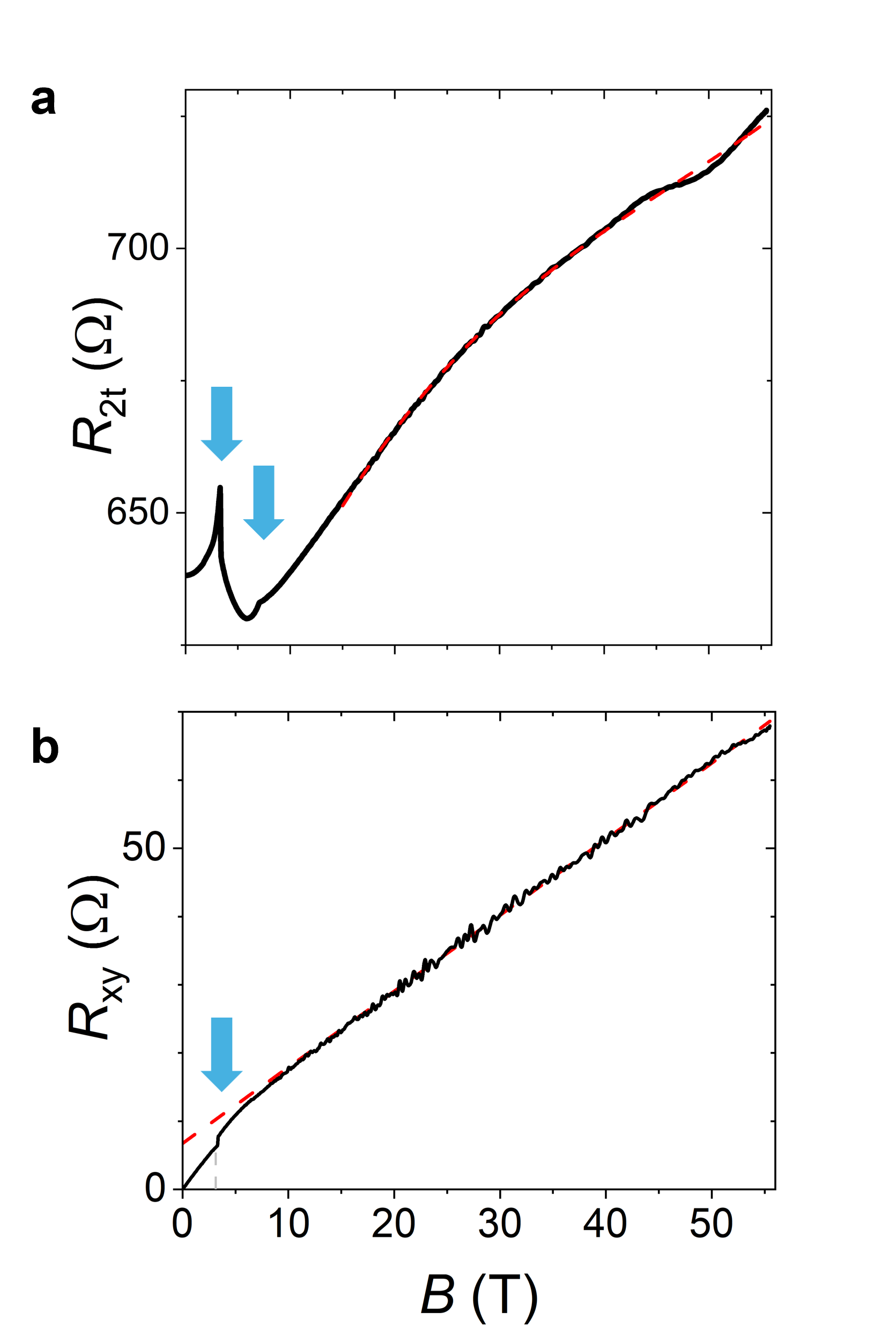}
\caption{
\textbf{Device MBT1 : Magneto-resistance and Hall response in the out-of-plane configuration.} The longitudinal magnetoresistance \textbf{(a)} and the transverse magnetoresistance \textbf{(b)} were measured at 4.2K up to 55T, and the data were symmetrized and antisymmetrized with respect to the magnetic field, respectively. Shubnikov-de-Haas oscillations are visible above about 40~T. Purple dashed lines represent respectively a cubic fit of the magnetoresistance, used afterward to extract Shubnikov-de-Haas oscillations, and a linear fit of the Hall effect asymptote at high field. Arrows indicate the spin-flop- and saturation fields.
}
\label{Fig:high-field}
\end{figure}

\begin{figure*}[htp]
\centering
\includegraphics[width=1\textwidth]{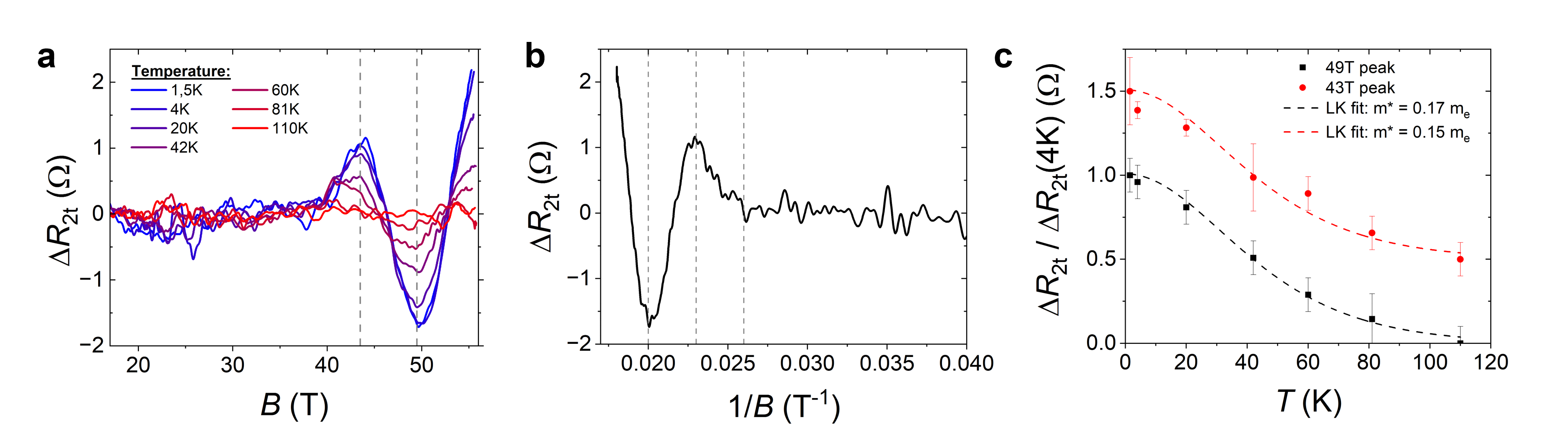}
\caption{ \textbf{Evidence of Shubnikov-de-Haas oscillations and Lifshitz-Kosevich analysis. a:} Residual Shubnikov-de-Haas oscillations after removal of a cubic background, at different temperatures. The position of the minimum and maximum of the oscillation are indicated by dashed lines.  \textbf{b:} Residual Shubnikov-de-Haas oscillations after removal of a cubic background, versus inverse magnetic field. The dashed lines show the position of the SdHO extrema, separated by 0.003~T$^{-1}$. \textbf{c:} Temperature dependence of the Shubnikov-de-Haas oscillation amplitude for the peaks at 43 and 49T. The error bars correspond to rapid noise in the raw data. \LK fits lead effective masses $m^*$ of about $0.15 - 0.17 m_\mathrm{e}$.
}
\label{Fig:T-dep}
\end{figure*}

Fig.\ref{Fig:high-field} presents the longitudinal and the transverse magnetoresistance under perpendicular magnetic fields up to 55~T. At low field, both samples show the expected signature of a collinear antiferromagnet with a uniaxial anisotropy: a sharp resistance peak at about 3.1~T corresponds to the spin-flop magnetic transition,\cite{Tan2020} which is also visible in the transverse resistance as a sharp jump of the anomalous Hall effect (blue arrows). Another feature is observed at 7~T, corresponding to the saturation field above which the magnetization is uniform (see Supplementary Information). This low-field magnetotransport, together with the observed Neel temperature, confirm that our nanostructures have all the usual magnetotransport properties expected of \MBT samples. Above 10~T, the transverse resistance is linear up to 55~T, corresponding to an asymptote of the normal Hall effect, yielding a total bulk carrier density of $6.7 \times 10^{19}$~cm$^{-3}$.

Above 40~T, oscillations appear in the longitudinal magnetoresistance. While only one and a half full oscillations are visible, those appear periodic in inverse magnetic field, as can be seen on Fig.\ref{Fig:T-dep}a. Their amplitude increases with the magnetic field, and they are damped by increasing temperature while remaining at the same position in field. We therefore identify them as Shubnikov-de-Haas oscillations. The appearance of SdHO at such high field reflects the moderate quantum mobility of the carriers, about $250 \, \text{cm}^2/\text{V.s} $. The periodicity in inverse magnetic field of 0.003~T$^{-1}$ corresponds to a magnetic frequency $f_\mathrm{B} = 167$~T, a parameter directly related to the subband carrier density. We investigate the temperature dependence of the oscillations after removal of a monotonic background for each temperature, as shown in Fig.\ref{Fig:T-dep}b. The oscillations are slowly damped by increasing temperature, until they disappear at around 110~K. From this temperature dependence, we extract the effective mass $m^*$ of the corresponding charge carriers through a \LK fit (see Fig.\ref{Fig:T-dep}c).\cite{Lifshitz1956} The obtained value is $m^* = 0.16 \pm 0.01 m_\mathrm{e}$, which could be compatible with both bulk states (close to $0.15 m_\mathrm{e}$: the effective mass of the second bulk conduction band CB2, but significantly different from the other two bulk bands, see below and \cite{Xu2021}) or with topological surface states with linear energy dispersion (for which $m^* = E_\mathrm{F} / v_\mathrm{F}^2$, with $E_\mathrm{F}$ and $v_\mathrm{F}$ the Fermi energy and Fermi velocity).\\

\begin{figure*}[htbp]
\centering
\includegraphics[width=1\textwidth]{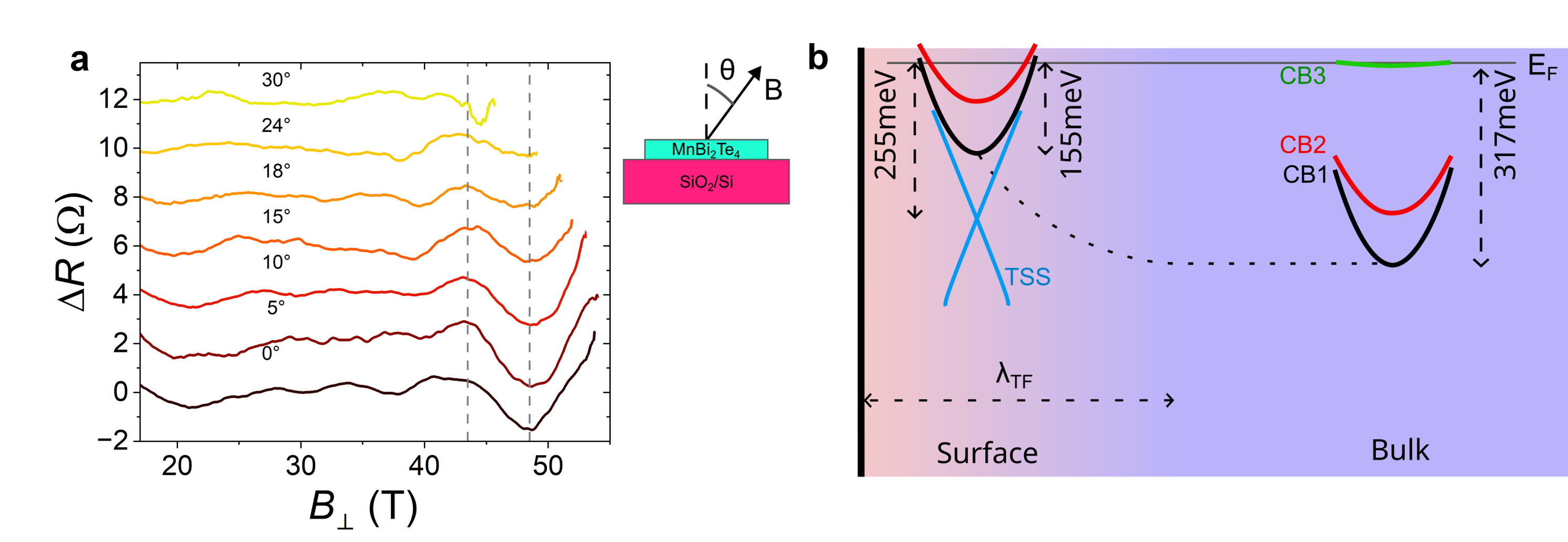}
\caption{ \textbf{Angular dependence of Shubnikov-de-Haas oscillations and band bending analysis. a:} Angular dependence of the Shubnikov-de-Haas oscillations after removal of a cubic background, between 0\textdegree~ (transverse field configuration) to 30\textdegree~ tilt toward in-plane configuration, shown against the transverse magnetic field component $B_\perp = B \cdot \mathrm{cos}(\theta)$. Curves shifted for clarity. The schematic shows the tilting angle $\theta$ of the magnetic field. \textbf{b:} Schematic of the energy levels and band bending in the \MBT nanostructure, as extracted from \SdH and Hall data (see main text), in a model with three conduction bands (CB1, CB2, CB3, see main text) and one topological surface state (TSS).
}
\label{Fig:theta-dep}
\end{figure*}

In order to determine the 3D or 2D character of the electronic population responsible for the observed SdHO, the magnetic field was tilted out of the sample plane by an angle $\theta$.  From the angular dependence of the magnetoresistance we observe that the SdHO evolves with the angle, but is stable with respect to $B_\perp = B\cos(\theta)$. Fig.\ref{Fig:theta-dep}a shows the SdHO plotted against the perpendicular component of the magnetic field $B_\perp$. The position of the maximum and minimum of the visible SdHO remains unaffected, while the positions vary with regard to the total field.
This is a clear sign that the SdHO originate from a 2D electronic state in the sample plane. This is also compatible with the carrier density extracted from the SdHO frequency: if interpreted as a 3D parabolic bulk band, a magnetic frequency $f_B = 167$~T corresponds to a bulk carrier density of $n_{\mathrm{3D}}^{\mathrm{SdHO}} = 1/2\pi^2 \, (2e/h \times f_\mathrm{B})^{3/2} = 8 \times 10^{17}$~cm$^{-3}$, which is two orders of magnitude smaller than the total carrier density extracted from the Hall effect. Interpreting the SdHO as originating from a topological surface state (TSS) with linear energy dispersion and a Fermi velocity $v_\mathrm{F} = 5.5 \times 10^5$~m.s$^{-1}$,\cite{Tomarchio2022,Gong2019} we calculate a 2D carrier density of $n_{\mathrm{2D}}^{\mathrm{SdHO}} = e/h \times f_\mathrm{B} = 4.1 \times 10^{12}$~cm$^{-2}$ at the topological surface state, and a Fermi energy $E_\mathrm{F}^\mathrm{TSS}=\hbar \cdot v_\mathrm{F}\cdot\sqrt{4\pi \cdot n_\mathrm{2D}^{\mathrm{SdHO}}}=$ 255~meV above the Dirac point in good agreement with ARPES \cite{Xu2024,Estyunin2020}.

In order to determine whether the surface state was localized on the top or the bottom surface of the nanostructure, we measured the dependence of the SdHO while applying a back-gate voltage (see Supplementary Information). As a large back-gate voltage of 110V did not induce any change on the SdHO, we conclude that the surface state responsible for the observed SdHO is not located on the bottom surface, but rather on the top one. The reason for the absence of visibility of SdHO originating from a bottom surface state probably lies in a lower mobility, which could be due to interaction with charge traps in the \ce{SiO2} substrate.

Consequently, we draw a model of the band filling and band bending in the nanostructure. This model, based on the density functional theory (DFT) calculations from \cite{Xu2021}, is summarized in Fig.\ref{Fig:theta-dep}b and detailed below. DFT predicts three conduction bands in the bulk centered on the $\Gamma$ point, CB1; CB2 and CB3, with respective effective masses $0.09 ~m_e$, $0.15 ~m_e$ and $3 ~m_e$, and bottom band energies at $80 ~\text{meV}$ and $310 ~\text{meV}$ below CB3 \cite{Xu2021}. We use this simplified model of the bulk band structure to calculate the energy profile of the nanostructure at the bulk and surface. In this simple calculation, the bands are supposed perfectly parabolic and isotropic.

Since the carrier density of a 91~nm thick nanostructure is likely to be dominated by the bulk, we calculate the bulk carrier density as the total carrier density extracted from high-field Hall slope: $n_{\mathrm{3D}}^{\mathrm{Hall}} = 6.7 \times 10^{19}$~cm$^{-3}$. In the 3-bulk-band model, this corresponds to a bulk Fermi level about 317~meV above the bottom of the conduction band CB1 (see Supplementary Information). The bulk chemical potential is therefore pinned at the bottom of the very heavy third conduction band CB3. While using a more exact model based on integrated density of states from DFT would change the exact value of the calculated bulk chemical potential, we argue that given the very large carrier density measured, the main result of the model - the very large bulk chemical potential - would not change qualitatively.

We now model the chemical potential at the surface using the charge carrier density extracted from the SdH oscillations. As calculated above, the chemical potential of the surface lies 255 meV above the Dirac point. Using a separation of 100meV between the Dirac point and the lowest conduction band CB1 from the literature (as determined for instance by ARPES, see \cite{Li2019f,Hao2019,Estyunin2020}), this imposes the position of the bulk band at the surface 155meV below the Fermi energy. As can be seen in Fig. \ref{Fig:theta-dep}, this results in an upward band bending of more than 150 meV between bulk and surface.
This upward band bending is a common situation in heavily doped 3DTI \cite{Veyrat2015}, where topological surface states are filled by charge transfer from heavily doped bulk bands, resulting in a lower doping of the bulk band close to the topological surface over the length-scale of the Thomas-Fermi screening length. It is, however, interesting to notice that, given the important band bending between TSS and bulk states, surface-sensitive techniques such as ARPES measurements would significantly underestimate the bulk chemical potential, and in particular the filling of the third conduction band CB3. Given the large effective mass of CB3, the bulk chemical potential will efficiently be pinned at the bottom of CB3 for carrier densities larger than $\sim 5\times 10^{19} \text{cm}^{-3}$.

We can further check the self-consistency of our model by calculating the effective mass expected for such topological surface states with such a Fermi energy. Using $m^* = E_\mathrm{F} / v_\mathrm{F}^2$, we obtain $m^* = 0.148 m_\mathrm{e}$, which is very close to the value extracted from the SdHO temperature dependence. This further confirms the 2D character of our SdHO. For TSS, the electronic mean free path is $l_\mathrm{e} = \mu \, m^* \, v_\mathrm{F}/e = \mu \, E_\mathrm{F} \, / \, e \, v_\mathrm{F}$. Considering the quantum mobility $\mu = 1/(40\, \text{T}) = 250 \, \text{cm}^{2}/\text{V.s} $ from the SdHO onset, this gives a lower estimate of the electronic mean free path $l_\mathrm{e} \simeq 11 \, \text{nm}$. This is lower than the value measured in non-magnetic TIs like Bi$_2$Se$_3$ (21nm, \cite{Dufouleur2016}), which is consistent with the higher doping level of \MBT as well as with the stronger chemical disorder, as due to cation intermixing. 

Summing up the results of analysis and discussion, the hypothesis of 2D surface state-origin SdH oscillations is supported by:\\
a) the 1/cos angle dependence of the peak position in magnetic field,\\
b) the effective masses $m^*$ obtained from the model being consistent with the ones extracted from the Lifshitz-Kosevich fit of the measured SdH oscillations.
This band bending model as obtained from the 3-band DFT model under the 2D hypothesis is further consistent with band bending reported on other topological insulators \cite{Veyrat2015}, and the surface chemical potential estimated from $f_\mathrm{B}$ is also consistent with ARPES results \cite{Xu2024, Estyunin2020}.

\smallskip

In conclusion, we have studied magnetotransport up to very high magnetic fields in Hallbars of exfoliated \MBTT. Above 40~T we evidence clear \SdH oscillations, due to the moderate mobility of the charge carriers in \MBTT. The angular dependence of the \SdH oscillations in tilted field reveals their 2D origin. From \SdH and Hall effect analysis we construct a model of band bending in our nanostructure, yielding properties of both bulk- and surface states. Our study presents the first transport evidence of surface states in \MBTT, which should play an important role in determining the transport properties in thin nanostructures. In particular, understanding and controlling the properties of the topological surface states is important to optimize the quantum anomalous Hall effect in magnetic topological insulators.\\
\\

\paragraph{\bf Supporting Information} 
Additional magnetotransport data on both \MBT devices, including gate-, angle- and field dependencies (PDF)

\paragraph{\bf Acknowledgements} 
This work was supported by the Deutsche Forschungsgemeinschaft (DFG, German Research Foundation) under Germany's Excellence Strategy through the W\"{u}rzburg-Dresden Cluster of Excellence on Complexity and Topology in Quantum Matter -- \emph{ct.qmat} (EXC 2147, 0242021).
LV was supported and by the French ANR (project number ANR-23-CPJ1-0158-01 and ANR-24-CE92-0021-01).
L.V. and J. D. were supported by the Leibniz Association through the Leibniz Competition.
This work was supported by the European Union’s H2020 FET Proactive project TOCHA (No. 824140), as well as by the CNRS International Research Project “CITRON”. We acknowledge the support of LNCMI-CNRS, a
member of the European Magnetic Field Laboratory (EMFL) under proposal TMA01-222.
\section*{References}

\bibliography{SdH}

\clearpage
\newpage
\begin{figure}
    \centering
    \includegraphics[width=1\textwidth]{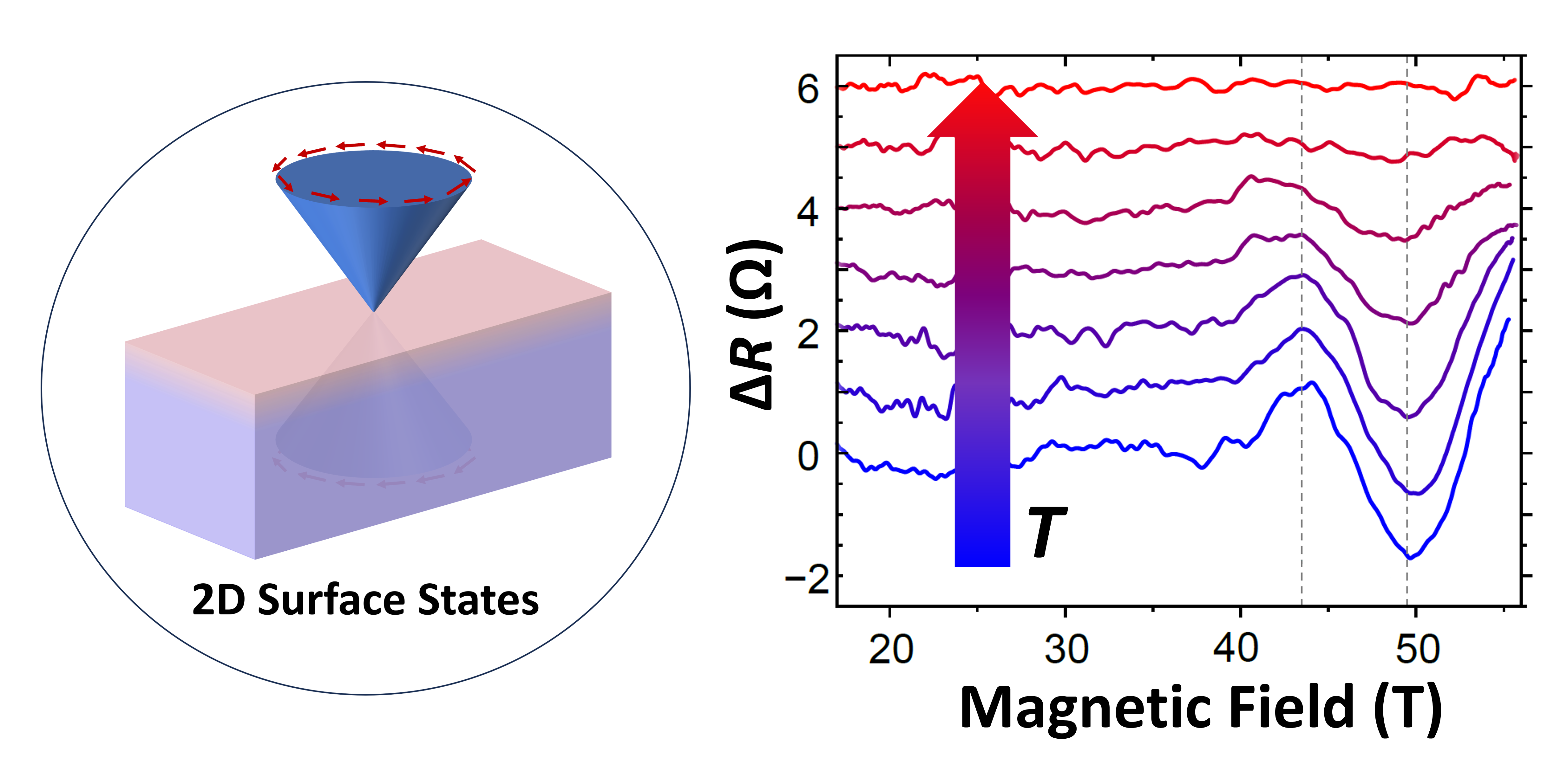}
\end{figure}

\end{document}